\documentclass[twocolumn,twocolappendix]{aastex631}
\usepackage{CJKutf8}
\usepackage{color}

\newcommand{\cifull}{\mbox{\rm [\ion{C}{1}] $^3P_1\text{--}^3P_0$}}
\newcommand{\ci}{\mbox{\rm [\ion{C}{1}]($1\text{--}0$)}}
\newcommand{\cishort}{\mbox{\rm [\ion{C}{1}]}}
\newcommand{\cishortit}{\mbox{\it [\ion{C}{1}]}}
\newcommand{\coone}{\mbox{\rm CO($1\text{--}0$)}}
\newcommand{\cojone}{\mbox{\rm CO($J=1\text{--}0$)}}
\newcommand{\siii}{\mbox{\rm [\ion{S}{3}]}}
\newcommand{\sii}{\mbox{\rm [\ion{S}{2}]}}

\newcommand{\oiii}{\mbox{\rm [\ion{O}{3}]}}
\newcommand{\kms}{\mbox{km s$^{-1}$}}
\newcommand{\Kkms}{\mbox{K km s$^{-1}$}}
\newcommand{\ergs}{\mbox{erg s$^{-1}$}}
\newcommand{\MPIA}{\affil{Max-Planck-Institut f\"{u}r Astronomie, K\"{o}nigstuhl 17, D-69117, Heidelberg, Germany}}
\newcommand{\Nichidai}{\affil{Department of Physics, General Studies, College of Engineering, Nihon University, 1 Nakagawara, Tokusada, Tamuramachi, Koriyama, Fukushima, 963-8642, Japan}}
\newcommand{\NAOJ}{\affil{National Astronomical Observatory of Japan, 2-21-1 Osawa, Mitaka, Tokyo, 181-8588, Japan}}
\newcommand{\IoA}{\affil{Institute of Astronomy, School of Science, The University of Tokyo, 2-21-1, Osawa, Mitaka, Tokyo 181-0015, Japan}}
\newcommand{\UFukushima}{\affil{Division of Human Support System, Faculty of Symbiotic Systems Science, Fukushima University, Fukushima 960-1296, Japan}}
\newcommand{\UVirginia}{\affil{Departments of Chemistry and Astronomy, University of Virginia, Charlottesville, VA 22904, USA}}
\newcommand{\SOKENDAI}{\affil{Department of Astronomy, School of Science, The Graduate University for Advanced Studies (SOKENDAI), 2-21-1 Osawa, Mitaka, Tokyo, 181-1855 Japan}}
\newcommand{\ISEE}{\affil{Institute for Space-Earth Environmental Research, Nagoya University, Furo-cho, Chikusa-ku, Nagoya, Aichi 464-8601, Japan}}
\newcommand{\UNagoya}{\affil{Division of Particle and Astrophysical Science, Graduate School of Science, Nagoya University, Furocho, Chikusa-ku, Nagoya, Aichi 464-8602, Japan}}
\newcommand{\UJoetsu}{\affil{Joetsu University of Education, Yamayashiki-machi, Joetsu, Niigata 943-8512, Japan}}
\newcommand{\RESCEU}{\affil{Research Center for the Early Universe, School of Science, The University of Tokyo, 7-3-1 Hongo, Bunkyo-ku, Tokyo 113-0033, Japan}}
\newcommand{\MPE}{\affil{Max-Planck-Institut f\"ur Extraterrestrische Physik (MPE), Giessenbachstr. 1, D-85748 Garching, Germany}}

\received{December 17, 2021}
\revised{February 23, 2022}
\accepted{March 2, 2022}

\shorttitle{Neutral Atomic Carbon Outflow in NGC~1068}
\shortauthors{T. Saito et al.}

\graphicspath{{./}{figures/}}

\begin{document}
\title{The Kiloparsec-scale Neutral Atomic Carbon Outflow in the Nearby Type-2 Seyfert Galaxy NGC~1068: Evidence for Negative AGN Feedback}

\correspondingauthor{Toshiki Saito}
\email{toshiki.saito@nao.ac.jp, saito.toshiki@nihon-u.ac.jp}

\author[0000-0002-2501-9328]{Toshiki~Saito}\Nichidai\NAOJ
\author[0000-0001-6788-7230]{Shuro~Takano}\Nichidai
\author[0000-0002-6824-6627]{Nanase~Harada}\NAOJ\SOKENDAI
\author[0000-0002-8467-5691]{Taku~Nakajima}\ISEE
\author[0000-0002-3933-7677]{Eva~Schinnerer}\MPIA
\author[0000-0001-9773-7479]{Daizhong~Liu}\MPE
\author[0000-0002-9695-6183]{Akio~Taniguchi}\UNagoya
\author[0000-0001-9452-0813]{Takuma~Izumi}\NAOJ\SOKENDAI
\author{Yumi~Watanabe}\UFukushima
\author[0000-0001-9720-8817]{Kazuharu~Bamba}\UFukushima
\author[0000-0002-4649-2536]{Eric~Herbst}\UVirginia
\author[0000-0002-4052-2394]{Kotaro~Kohno}\IoA\RESCEU
\author[0000-0003-0563-067X]{Yuri~Nishimura}\IoA\NAOJ
\author[0000-0002-9333-387X]{Sophia~Stuber}\MPIA
\author[0000-0003-4807-8117]{Yoichi~Tamura}\UNagoya
\author[0000-0001-9016-2641]{Tomoka~Tosaki}\UJoetsu

\begin{abstract}
Active galactic nucleus (AGN) feedback is postulated as a key mechanism for regulating star formation within galaxies. Studying the physical properties of the outflowing gas from AGN is thus crucial for understanding the co-evolution of galaxies and supermassive black holes. Here we report 55~pc resolution ALMA neutral atomic carbon \cifull\ observations toward the central 1 kpc of the nearby type-2 Seyfert galaxy NGC~1068, supplemented by 55~pc resolution \cojone\ observations. We find that \cishort\ emission within the central kpc is strongly enhanced by a factor of $>$5 compared to the typical \cishort/CO intensity ratio of $\sim$0.2 for nearby starburst galaxies (in units of brightness temperature). The most \cishort-enhanced gas (ratio $>$ 1) exhibits a kpc-scale elongated structure centered at the AGN that matches the known biconical ionized gas outflow entraining molecular gas in the disk. A truncated, decelerating bicone model explains well the kinematics of the elongated structure, indicating that the \cishort\ enhancement is predominantly driven by the interaction between the ISM in the disk and the highly inclined ionized gas outflow (which is likely driven by the radio jet). Our results strongly favor the ``CO dissociation scenario" rather than the ``in-situ C formation" one which prefers a perfect bicone geometry. We suggest that the high \cishort/CO intensity ratio gas in NGC~1068 directly traces ISM in the disk that is currently dissociated and entrained by the jet and the outflow, i.e., the ``negative" effect of the AGN feedback.
\end{abstract}


\section{Introduction} \label{sec:intro}
Galactic outflows are a key phenomenon and feedback mechanism in galaxies regulating the star formation and galaxy quenching. These can be driven by active star formation, accreting supermassive black holes (active galactic nuclei; AGN), or both at the centers of galaxies. Especially, cold gas outflows are a subject of great interest, as they contain the raw material from which stars are formed, and thus potentially destine the evolution of galaxies \citep[see review by][]{Veilleux20}. AGN-driven outflows affecting the surrounding interstellar medium (ISM), i.e., AGN feedback, are considered to play a critical role in the co-evolution of supermassive black holes and their host galaxies \citep[e.g.,][]{Magorrian98,Costa14}.

However, there are two long-standing, contradictory paradigms regarding the effect of AGN feedback: ``positive" feedback and ``negative" feedback. In the positive feedback scenario AGN outflows or jets induce star formation through the interaction and compression of the surrounding materials \citep[e.g.,][]{vanBreugel93,Silk13}.
Negative feedback, in contrast, suppresses or even inhibits star formation by sweeping up and heating ISM \citep[e.g.,][]{Croft06,Sturm11}. Understanding the exact AGN feedback mechanism and its role in galaxy evolution is thus one of the challenges of modern astrophysics.

In this Letter, we present tangible evidence for ongoing negative feedback in the nearby ($\sim$13.97~Mpc; \citealt{Anand21}) type-2 Seyfert galaxy NGC~1068 using high-quality ALMA \cifull\ (hereafter \ci) and \coone\ datasets. The outflow and the jet of NGC~1068 are well-studied because of their close distance and apparent brightness \citep[e.g.,][]{Wilson83,Das06,Barbosa14,Garcia-Burillo14,Garcia-Burillo19}. Based on our finding as well as multi-wavelength studies in the literature, we provide a plausible explanation on what drives and how significant is the negative effect of the AGN feedback in NGC~1068.

\begin{figure*}[t!]
\begin{center}
\includegraphics[width=18cm]{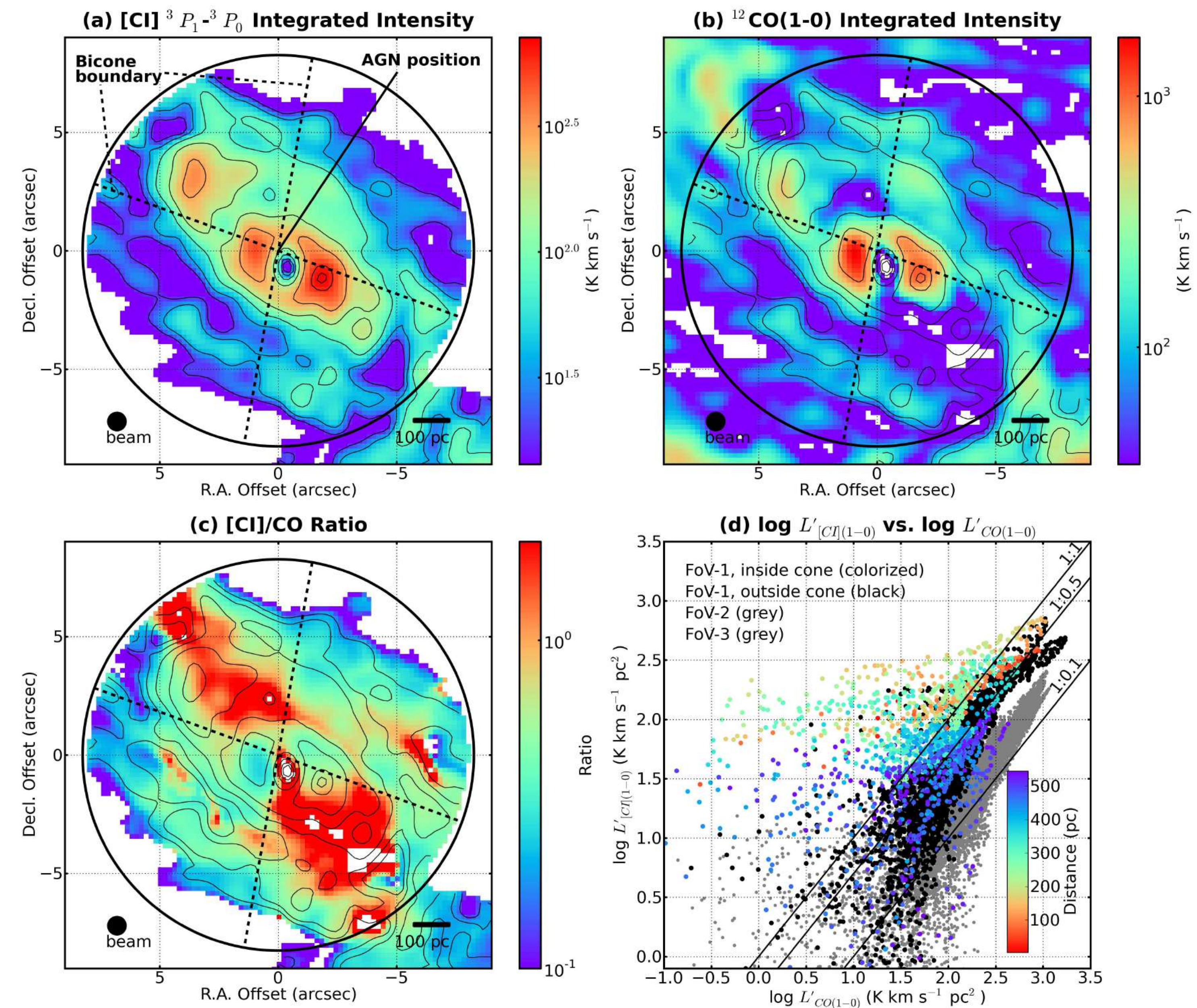}
\end{center}
\caption{
(a) \ci\ integrated intensity map of the central 1-kpc of NGC~1068 (FoV-1) at 0\farcs8 resolution ($\sim$ 55~pc). The contours are (0.025, 0.05, 0.1, 0.2, 0.4, 0.8, and 0.96) $\times$ peak, where peak $=$ 714~\Kkms. The two dashed lines crossing the AGN position \citep{Roy98} denote the approximate outer edges of the ionized gas cones \citep{Mingozzi19}. The black circle indicates the FoV of the Band 8 12m+7m data.
(b) \coone\ integrated intensity map at 0\farcs8 resolution (Tosaki et al. in prep.). The \ci\ contours are overlaid.
(c) \ci/\coone\ integrated intensity ratio map.
(d) Pixel-by-pixel comparison between the \ci\ and \coone\ maps. Pixel size is 0\farcs2 $\times$ 0\farcs2. Pixels within the cones in FoV 1 are colorized by their projected distance from the AGN position, and pixels outside the cones in FoV 1 are shown in black. Pixels in FoV 2 and FoV 3 (two positions in the starburst ring; Takano et al. in preparation) are shown in grey for reference.
\label{fig:map}}
\end{figure*}

\section{Observations and Data Processing} \label{sec:data}
{\it ALMA data summary:} We observed \ci\ at $\nu_{\rm rest}$ $=$ 492.16065~GHz using Band~8 and \coone\ at $\nu_{\rm rest}$ $=$ 115.27120~GHz using Band~3 (2017.1.00586.S; S. Takano et al. in preparation, and 2018.1.01684.S; T. Tosaki et al. in preparation) using the ALMA 12m array and the ACA 7m array. The 12m+7m \ci\ data (\coone\ data) achieved a spatial resolution of $\sim$0\farcs7 ($\sim$0\farcs4) and a maximum recoverable scale (MRS) of $\sim$14\arcsec ($\sim$59\arcsec) with native spectral resolution of 1.19~\kms\ (0.74~\kms). The achieved noise rms levels are $\sim$0.05~K for the \ci\ cube and $\sim$0.35~K for the \coone\ cube.

As the \ci\ field of view (FoV) ($\sim$16\arcsec) is much smaller than the \coone\ FoV ($\sim$71\arcsec), we targeted \cishort\ line emission from three representative positions of NGC~1068, i.e., the circumnuclear disk (CND) and the bar (FoV-1) and the southern and eastern parts of the starburst ring (FoV-2 and FoV-3). In this Letter, we focus on the central 1-kpc FoV-1, and the details of the remaining data will be presented by Takano et al. (in preparation).

Both our \ci\ and CO data might suffer from interferometric spatial filtering effect. As the MRS of the \ci\ data is smaller than that of the CO data, the \cishort/CO line ratios presented in this Letter can be regarded as the lower limits. Thus, this does not affect our discussion and conclusion on the extremely high \cishort/CO ratio gas. In addition, the missing flux effect might be minor or negligible, as the spatial structures of the \cishort/CO line ratio gas are basically smaller than the MRS of the \ci\ data ($\sim$14\arcsec).


{\it Imaging:} We performed the observatory-delivered calibration with minor manual data flagging (e.g., baseline and time flagging). Then, we reconstructed images using the {\tt PHANGS-ALMA imaging pipeline} \citep{Leroy21a}. During the imaging process of our data, single-scale {\tt CLEAN} was only employed (i.e., no multi-scale {\tt CLEAN}). Besides this, we followed the recommended standard setups (see \citealt{Leroy21a} for more details).

{\it Post-processing:} After imaging, the clean products are corrected for primary beam, convolved to a round 0\farcs8 beam ($\sim$55~pc), and regridded to 0\farcs2 pixel size. Then, moment maps are extracted from the data cubes using the imaging pipeline. As we focus on line ratios in this study, high S/N in both lines is required. Therefore, we use the ``strictly masked" moment map, which is based on a signal-to-noise ratio (SNR) mask (see \citealt{Leroy21a}), unless otherwise noted. The mask basically consists of all voxels with SNR $>$ 4 over two successive velocity channels, and then expanded to the contiguous voxels with SNR $>$ 2. The ``strict" moment maps are characterized to have high confidence but exclude faint structures, i.e., have lower completeness.

\section{Results} \label{sec:results}
We show the 55~pc resolution \ci\ and \coone\ integrated intensity maps toward the central 1 kpc of NGC~1068 in Figures~\ref{fig:map}a-b. As observed in many other galaxies (e.g., \citealt{Miyamoto21} and references therein), \ci\ basically traces structures seen in \coone\ (but see also \citealt{Michiyama20} for an exceptional case). However, if we focus on the projected area of the biconical ionized gas outflow \citep[e.g.,][hereafter D06]{Das06}, whose outline is highlighted by two dashed lines, the \cishort\ distribution is more extended than CO.

\subsection{\texorpdfstring{\cishort}{CI} vs. CO at 55 pc scale} \label{sec:ci_vs_co}
In order to quantify the apparent differences, we compare the two maps on pixel-by-pixel basis (Figures~\ref{fig:map}c-d). The log--log scatter plot and the \cishort/CO line ratio map show three characteristic features:

\begin{figure*}[ht!]
\begin{center}
\includegraphics[width=8.5cm]{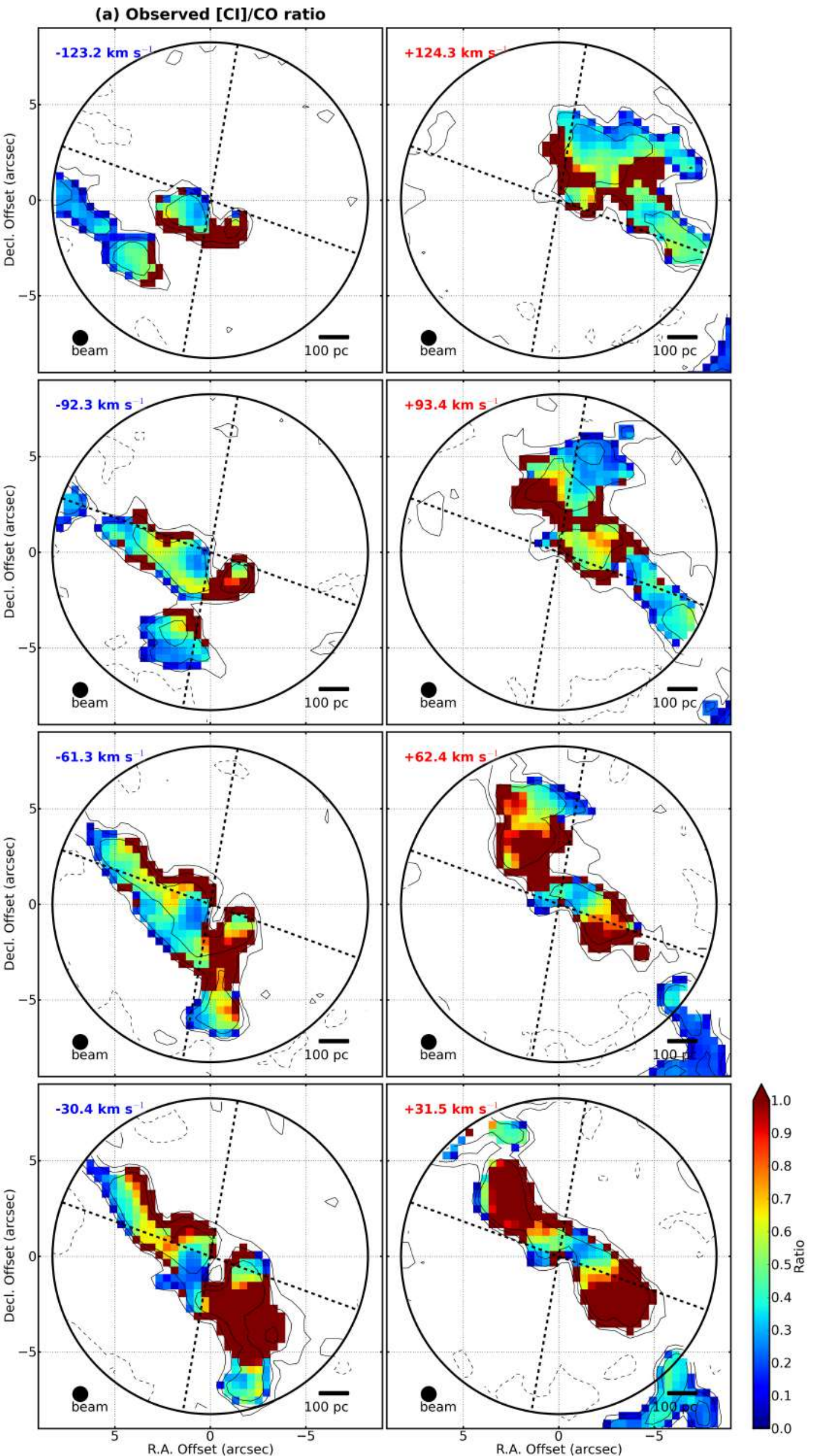}
\includegraphics[width=8.5cm]{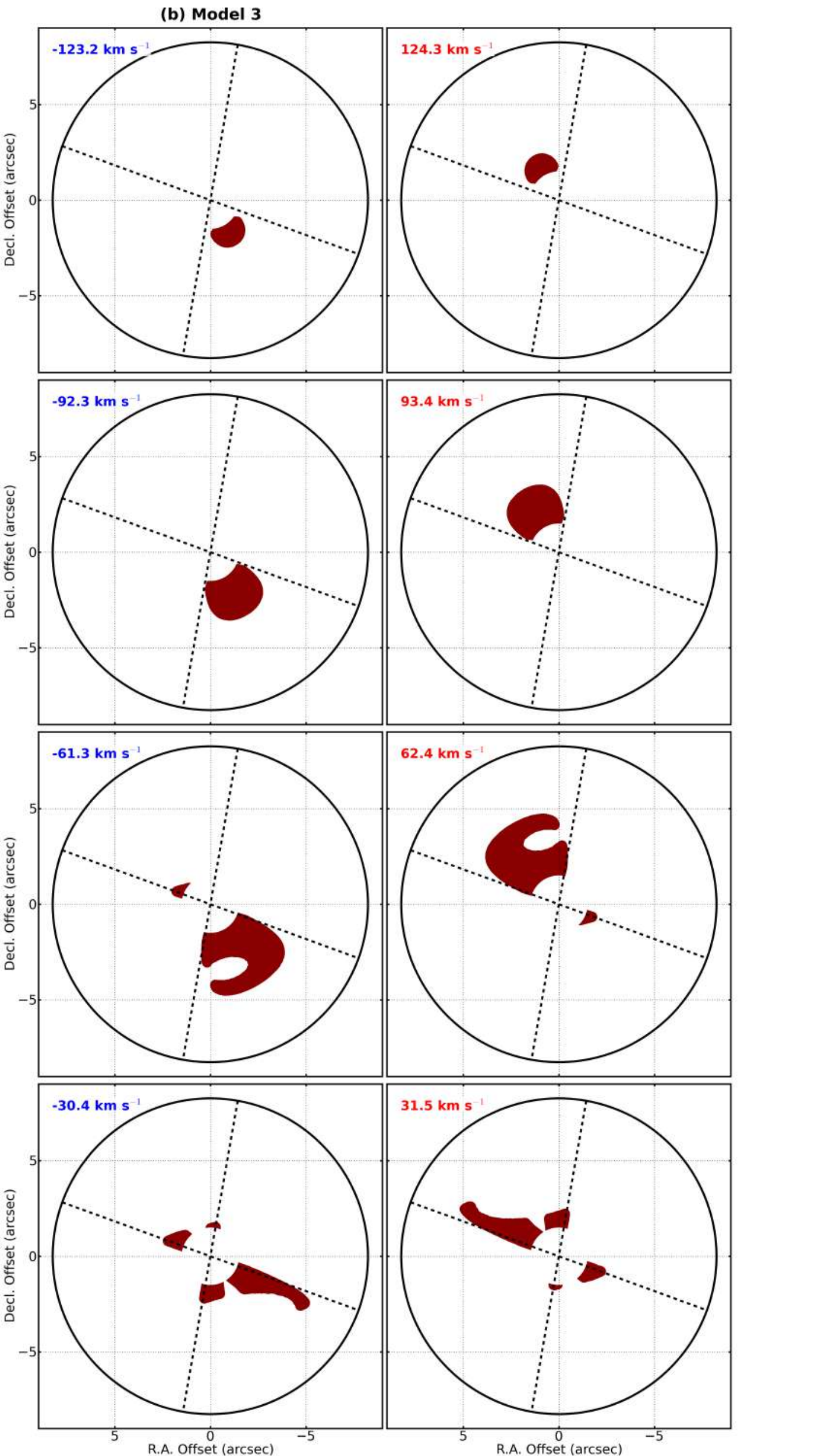}
\end{center}
\caption{
(a) \cishort/CO intensity ratio channel maps of NGC~1068. We show pixels with a lower limit of the ratio higher than unity in the reddest color (i.e., CO non-detected pixels but detected in \cishort). The left (right) panels show blueshifted (redshifted) part from the systemic velocity. The relative velocity offset is larger in the upper panels. The \cishort\ contours are overlaid.
(b) Channel maps of the truncated, decelerating bicone model. This simple model largely reproduces the kinematics of the observed high ratio gas shown in red.
\label{fig:channel}}
\end{figure*}

\begin{itemize}
\item {\it Highest ratios:} Pixels coinciding with the biconical ionized gas outflow (FoV-1; colorized points in Figure~\ref{fig:map}d) show \cishort\ emission comparably bright to CO in Kelvin units. The highest ratios are located 200-300~pc away from the AGN position (Figure~\ref{fig:map}c), which is {\it beyond} the outer radius of the CND ($r_{\rm out}$ $\simeq$ 200~pc; e.g., \citealt{Garcia-Burillo14}). The 5$^{th}$-16$^{th}$-50$^{th}$-84$^{th}$-95$^{th}$ percentiles of the ratio distribution are 0.16-0.33-0.72-1.86-4.86.
\item {\it Moderate ratios:} Pixels not affected by the ionized gas outflow in FoV-1 (black points) show moderately high line ratios. The percentiles are 0.12-0.22-0.39-0.62-0.86. This is similar to the \cishort/CO line ratio of 0.5-0.9 toward the AGN position of the Seyfert galaxy NGC~7469 \citep{Izumi20}. In NGC~7469, the highest line ratio is found in the vicinity of the AGN position at 100~pc resolution. This is quite different from the NGC~1068 case, where line ratios ($>$0.5) are high almost everywhere in the central kpc region and the highest ratio is outside the CND.
\item {\it Low ratios:} The scatter plot shows a superlinear relation for the starburst ring (FoV-2 and FoV-3; grey points). In stark contrast to FoV-1, \replaced{where}{almost} all the data points show \cishort/CO ratio $<$ 0.5. The five percentiles are 0.04-0.08-0.15-0.23-0.45. The superlinear \cishort--CO relation and the lower ratio are common features found in nearby starburst galaxies and their nuclei \citep[e.g.,][]{Jiao19,Salak19,Saito20}.
\end{itemize}

Further investigation of the different \cishort\ properties among FoV-1, FoV-2, and FoV-3 is beyond the scope of this dedicated outflow study, and will be discussed in a forthcoming paper. We note that we applied the same scatter analysis to the datacubes before collapsing and found exactly the same trend as described above.

\subsection{The kinematics of the high \texorpdfstring{\cishortit/CO}{Ci/CO} ratio gas} \label{sec:kinematics}
The \cishort/CO integrated intensity ratio map shows unusual line ratio values as described above, although this is not enough to fully uncover the characteristics of the \cishort/CO line ratio in NGC~1068. In this section, we try to constrain the geometry and kinematics of the high ratio gas.

In Figure~\ref{fig:channel}a, we show channel maps of the \cishort/CO intensity ratio within FoV-1. Each datacube is rebinned to 2.38~\kms\ spectral resolution before calculating the ratio. Then, the systemic velocity of NGC~1068 ($v_{\rm sys}$ $=$ 1116~\kms; \citealt{Garcia-Burillo14}) is subtracted. Here we follow the radio velocity convention with the kinematic LSR frame. In these channel maps, in addition to the pixels detected in both the \cishort\ and CO lines, we show pixels with significant \cishort\ detection but CO non-detection in the reddest color (i.e., pixels with a ratio higher than unity).

In each channel, the highest ratio gas shows the same trend as seen in the integrated intensity ratio map (see Section~\ref{sec:ci_vs_co}), i.e., highest ratio pixels ($>$1) preferentially appear within the ionized gas bicone whereas moderate ratios are outside the bicone. The highest ratio gas shows a systematic velocity structure with extended, low-velocity components ($|v-v_{\rm sys}|$ $\lesssim$ 65~\kms) and compact, high-velocity components ($\gtrsim$80~\kms). This systematic motion seems to deviate from the nearly east-west velocity gradient due to galactic rotation which is well traced in the low ratio gas (green-to-blue pixels in Figure~\ref{fig:channel}a; see \citealt{Schinnerer00} for a detailed modeling of the disk kinematics).

\subsection{Modeling the channel map} \label{sec:modeling}
Motivated by the properties of the \cishort\ line described above and the similarity between the high ratio gas distribution and the projected distribution of the ionized gas bicone, here we compare the observed channel map with channel maps reproduced from simple bicone models.

In order to create model channel maps, we follow the bicone modeling method described in D06. They reproduced the results of multiple long-slit observations of the \oiii\ line around the center of NGC~1068 using a simple bicone model, which is widely accepted and consistent with the results based on different methods and different datasets \citep[e.g.,][]{Barbosa14}.

The position angle, (outer) opening angle, and inclination of our bicone model are exactly the same as the D06 model, i.e., 30\degr, 40\degr, and 5\degr (90\degr\ for pole-on), respectively. The bicone geometry on the sky is, in short, nearly face-on and the north-east cone is a bit closer. The maximum distance from the nucleus along the bicone axis is set to 400~pc as D06 did. This bicone has an inner opening angle of 20\degr, so that it is hollow but has a certain thickness defined by the outer and inner opening angles. After assigning a certain outflow velocity structure to the bicone, we calculate, for each element composing the bicone in the $xyz$ space, its projected position and velocity on the sky. From this projected model we extract channel maps matching our data. After convolving the model with a 0\farcs8 beam, we mask the central 4\farcs0 diameter as the observed data are severely affected by the CND. Since the \oiii\ outflow velocity assigned to the D06 model (intrinsic velocity $v_{\rm int}$ $\lesssim$ 2000~\kms\ and projected velocity $v_{\rm proj}$ $\lesssim$ 1000~\kms) is much faster than the observed $v_{\rm proj}$ of the \cishort\ outflow ($\lesssim$150~\kms; Figure~\ref{fig:channel}a), we explore the velocity structure of the \ci\ outflow using three different models.

After careful exploration of the three models, we conclude that only the 3rd model can well reproduce the observed kinematics of the high \cishort/CO ratio gas. Here we briefly explain the three bicone models:
\begin{itemize}
\item {\it Model 1:} We start from a bicone model with a constant outflow velocity ($v_{\rm int}$ $=$ 300~\kms) as shown in Figure~\ref{fig:channel_model}a in Appendix~\ref{sec:appendix_a}. The nearly face-on geometry and the wide opening angle naturally result in the bicone shape in both the blue- and redshifted channels. The two striking failures of this model are (A) the presence of extended high-velocity components and (B) the presence of northern blueshifted and southern redshifted components, both of which are not seen in the real data. Note that varying $v_{\rm int}$ does not improve this model in terms of these aspects.

\item {\it Model 2:} In order to achieve compact distributions in high-velocity channels (failure A), we introduce a varying velocity structure to model 1. D06 required deceleration after acceleration in order to explain the \oiii\ observations. To do so, they introduced a turnover radius, $r_t$, 140~pc away from the nucleus. The outflow velocity peaks at $r_t$. We simply employ this. However, as $r_t$ is comparable to the CND size, the accelerating part is masked in our model channel maps. Therefore, we only model the decelerating part. We define $v_{\rm int}(r) = v_{\rm max}-k(r-r_t)$, where $v_{\rm max}$ is the maximum velocity of 300~\kms, $k$ is $v_{\rm max}/3r_t$, and $r$ is the distance from the nucleus ($v_{\rm int}$ reaches 0~\kms\ at $r = 4r_t$). The model channel maps are shown in Figure~\ref{fig:channel_model}b in Appendix~\ref{sec:appendix_a}. As intended, model 2 reproduced the compact distributions in high velocity channels. However, failure B of model 1 is still visible, and could not be solved with any choice of $k$ and $v_{\rm max}$.

\begin{figure*}[ht!]
\begin{center}
\includegraphics[width=16cm]{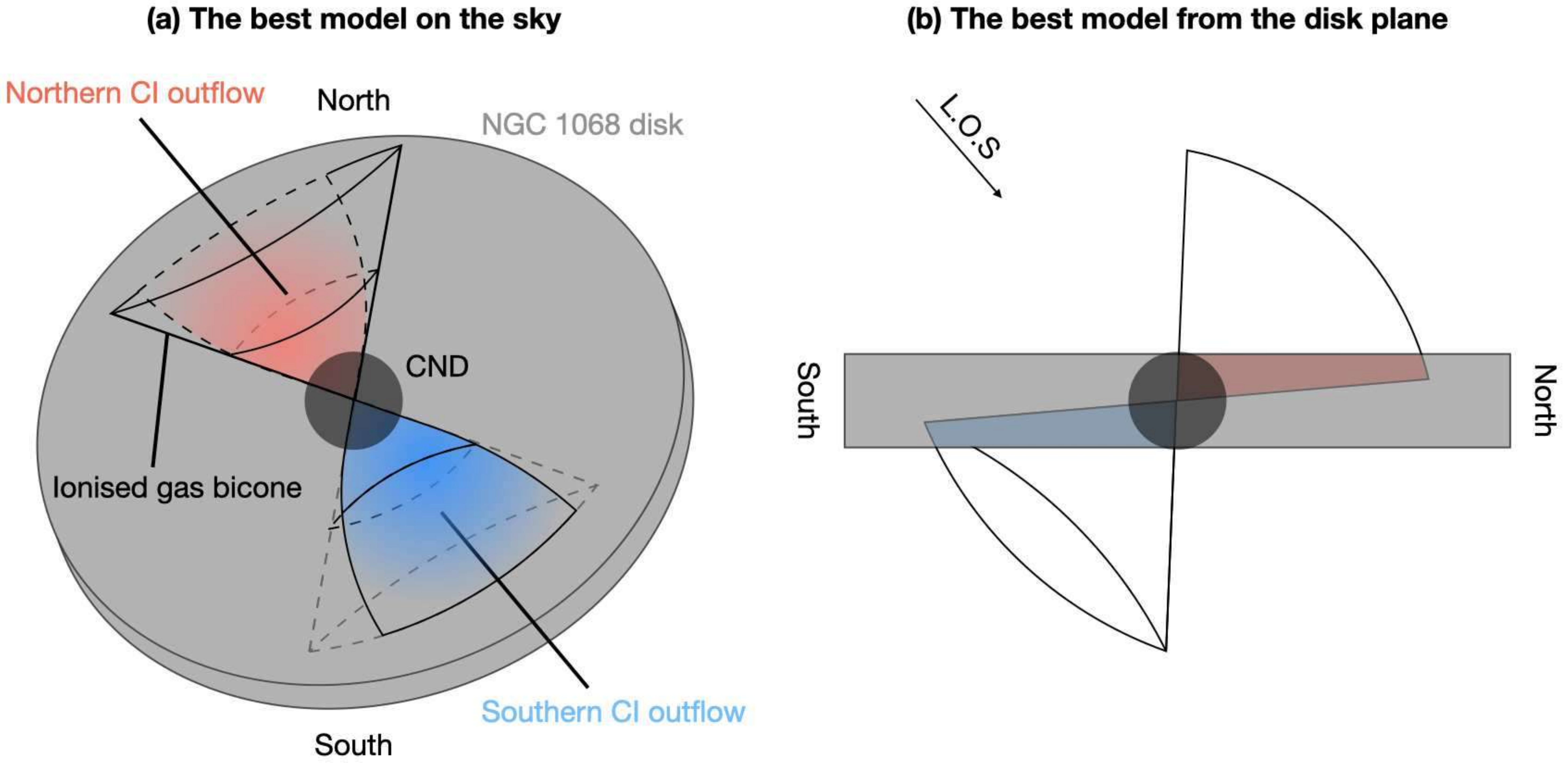}
\end{center}
\caption{
Illustrations of the truncated, decelerating bicone model (model 3). (a) View from Earth. (b) View from the disk plane. The high \cishort/CO ratio gas is distributed within the red and blue parts (i.e., overlap regions between the bicone and the disk).
\label{fig:3dmodel}}
\end{figure*}

\item {\it Model 3:} After some exercises with models 1 and 2, we come to a conclusion that a ``perfect" bicone geometry can never reproduce the observed channel map (failure B), because the nearly face-on geometry and the wide opening angle should always give both the blue- and redshifted components to both the northern and southern cones regardless of the velocity structure. More specifically, the foreground part ($=$ blueshifted) of the northern cone and the background ($=$ redshifted) part of the southern cone are not visible in the real data.

Recalling the D06 model, one important information that our models 1 and 2 are missing is the possible interaction with the galactic molecular gas disk. This is also clearly seen in the schematic illustration provided by \citet{Garcia-Burillo19}. This interaction is actually expected considering the geometry of the nearly face-on bicone ($i =$ 5\degr) and the relatively face-on galactic molecular gas disk \citep[$i =$ 41\degr;][]{Schinnerer00,Garcia-Burillo14}. Here we reproduce the ``imperfect" (or truncated) bicone geometry based on the assumption that high \cishort/CO line ratio gas is arising from regions where the bicone overlaps with the galaxy disk in $xyz$ space. We assume a constant height of the molecular gas disk of 150~pc, a disk position angle of 278\degr, and a disk inclination of 41\degr. The disk thickness is consistent with the scale heights measured for nearby spiral galaxies \citep[50-200~pc;][]{Patra19}. The resultant channel maps are shown in Figure~\ref{fig:channel}b. Although some substructures in the low velocity channels are not exactly recovered (which are likely due to the contribution from galaxy rotation and the presence of the bar), the overall trend seen in the high velocity channels of the real data is well reproduced.
\end{itemize}

Model 3 best reproduces the observed channel maps of the high \cishort/CO ratio gas with only three assumptions: (1) a bicone geometry exactly matching the known ionized gas bicone, (2) a decelerating velocity structure at $r > r_t$, and (3) an interaction with the galaxy disk (See Figure~\ref{fig:3dmodel} for illustrations of the orientation.). We assumed the geometry employed by the D06 \oiii\ outflow model; interestingly, the spatial configuration of model 3 is consistent with the molecular outflow model described by \citet{Garcia-Burillo14,Garcia-Burillo19}. This implies that the \oiii, \cishort, and CO outflows are closely related to each other.

\subsection{The \texorpdfstring{\cishort}{CI} bicone model in comparison with observations and simulations} \label{sec:comparison}
In the previous Section, we found that the spatial distribution of the high \cishort/CO ratio gas ($>$1) in NGC~1068 can be simply explained by the interaction between the hundred-pc-scale biconical outflow of the ionized gas (e.g., D06) and the molecular gas disk. Here we briefly discuss how this \cishort\ picture fits to previous observational and theoretical studies.

Many previous observational studies revealed that there is a tight spatial correlation between the AGN narrow-line region (e.g., D06; \citealt{Mingozzi19}), the radio jet \citep[e.g.,][]{Gallimore96,Capetti97}, and the entrained molecular gas in NGC~1068 \citep[e.g.,][]{Krips11,Garcia-Burillo14} from a few tens of pc up to $\sim$500~pc, all implying that the outflow and the jet in NGC~1068 (1) are driven by the central AGN and (2) interact with the CND and the galactic molecular gas disk (see also \citealt{Barbosa14}). Although the exact cause for the hundred-pc-scale ionized gas outflow is still under debate \citep[e.g., see][]{May17}, many models agree that the jet from the AGN (and its interaction with the multiphase ISM in the disk) is the origin of all the outflow phenomena in NGC~1068.

This jet-ISM interaction is observationally and theoretically known to produce an expanding energy bubble. This heats the surrounding medium and creates shocks at the colliding interface resulting in radial multiphase outflows \citep[e.g.,][]{Sutherland07,Matsushita07,Nesvadba08,Wagner11,Wagner12,Morganti15}. \citet{Mukherjee16} suggested that weak radio jets (radio power $\lesssim$ 10$^{43}$~\ergs) are less efficient in accelerating and sweeping clouds, but are able to affect the ISM over a large volume (especially in the lateral direction) because weak jets are trapped by the ISM for a long time. This ``weak" radio jet scenario is consistent with the estimated radio power of NGC~1068 ($1.8 \times 10^{43}$~\ergs; \citealt{Garcia-Burillo14}).

The low \cishort\ outflow velocity in our model ($v_{\rm max}$ $=$ 300~\kms) compared to the ionized gas outflow velocity ($v_{\rm max}$ $=$ 2000~\kms; D06) supports the idea that shocks propagating from the weak jet are insufficient to accelerate molecular clouds in the disk, but do efficiently heat the ISM as suggested by \citet{Mukherjee16}.

The shocks propagating through the dense gas heat, dissociate, and ionize the gas. The subsequent emission also produces ionizing and dissociating photons. We expect that those destructive processes, shock, dissociation, and ionization, happen in the overlap region between the outflow and the disk at the same time. Here we briefly discuss two possible mechanisms enhancing \cishort: shock-dissociation and photodissociation.

In the shocked gas, H atoms (shock products of efficient H$_2$ destruction) endothermically dissociate CO and produce C atoms \citep[e.g.,][]{Hollenbach80}.
\citet{Hollenbach80} also mentioned that the detection of high-speed interstellar molecules indicate that molecular gas reformed in the post-shock gas. However, if the reformation process dominates in the outflow, the \cishort/CO ratio should return to the original value. We take $\sim$0.2 as observed in FoV-2 and FoV-3 as representative for gas not affected by the outflow. Thus, although we expect that both shock dissociation and reformation happen in the post-shock gas, dissociation has to overwhelm reformation in order to explain the observed high \cishort/CO ratio. In addition, as suggested by \citet{Garcia-Burillo17}, the enhancement of C$_2$H in the molecular outflow of NGC~1068 is likely due to the jet-driven dissociation and short-lived ($\sim$10$^2$-10$^3$~yr$^{-1}$), implying that short-lived dissociation is an ongoing, continuous process in the outflow.

Similar to the shock dissociation as described above, when clouds are irradiated by strong UV field, CO column density decreases and C column density increases \citep{Meijerink05}. \citet{Meijerink05} also predicted an enhancement of C atoms due to X-rays. All those processes likely result in higher \cishort/CO intensity ratios. The presence of kpc-scale hot outflows in X-ray and FUV \citep[e.g.,][]{Ogle03} observationally supports this picture. However, based on the current data and analysis, we cannot rule out that UV and X-ray photons coming from the AGN play an important role in the molecular gas dissociation/ionization in the overlap region.


Combining our \cishort\ study with multi-wavelength observations (Figures~\ref{fig:multi}a-d) and simulations, we suggest that the highly-inclined jet (radiatively and mechanically) heats, dissociates, and ionizes the ISM in the central kpc of the disk (i.e., neutral gas clouds are not launched from the AGN). Thus, the AGN jet feedback suppresses ``future" star formation in this part of the disk by destroying molecular clouds, not by efficiently sweeping them. A part of the phenomenon is observed as a strong pixel-by-pixel correlation between our \cishort/CO ratio map (i.e., molecular cloud dissociation) and the \siii/\sii\ ratio map \citep[i.e., high ionization;][]{Mingozzi19} at 55~pc resolution (Pearson correlation coefficient $r$ = 0.62; see Figures~\ref{fig:multi}e).

One could consider that strong UV radiation from young massive stars forming in the clouds entrained in the outflow can explain the observed high \cishort/CO ratio, i.e., positive feedback scenario.
However, taking the $v_{\rm int}$ ($\sim$2000~\Kkms) and the maximum distance from the nucleus ($\sim$400~pc) of the ionized gas outflow (D06), the timescale of the outflow phenomenon is just $\sim$0.2~Myr. As this is too short to allow the clouds to form stars (typical star formation timescale is 30~Myr; e.g., \citealt{Kawamura09}), we deem the positive feedback scenario unlikely.

We note that there is an alternative way to realize multiphase outflows, which is an ``in-situ cold gas formation" scenario (e.g., see \citealt{Girichidis21} and references therein). However, this prefers a ``perfect" bicone geometry for the cold gas outflow as gas cooling isotropically happens, which is inconsistent with the observed truncated bicone. In addition, the efficient cooling may result in CO enhancement not only C, that is unlikely the case of NGC~1068, as the CO map shows holes toward the bicone direction (Figure~\ref{fig:map}b).

\section{Summary and Implications} \label{sec:summary}
In this Letter, we present high resolution (0\farcs8) ALMA observations of \ci\ and \coone\ toward the central 1-kpc of the type-2 Seyfert galaxy NGC~1068. We found a kpc-scale elongated structure with extremely high \cishort/CO intensity ratio ($>$1), which coincides well with the spatial distribution of the known biconical ionized gas outflow. Our simple kinematic modeling, combining with multi-wavelength studies in the literature, yields that the high ratio gas is likely due to interaction of the jet/ionized gas outflow with the galaxy disk. This interaction results in efficient dissociation (high \cishort/CO ratio) of molecular clouds within the disk plane. Thus, this is clear evidence for the negative AGN feedback in the central 1-kpc of NGC~1068. We suggest that the relatively weak radio jet results in the large-scale molecular gas dissociation which is more efficient than the sweeping effect at this stage.

Our current analysis is based on line intensities. This makes it difficult to constrain the physical and chemical properties of the negative AGN feedback happening in NGC~1068 by direct comparison with models. Thus, an accurate measurement of the \cishort\ column density by observing the other \cishort\ transition represents a future direction.

\begin{acknowledgments}
This work was supported by NAOJ ALMA Scientific Research Grant Numbers 2021-18A. NH acknowledges support from JSPS KAKENHI Grant Number JP21K03634. EH acknowledges support from the US National Science Foundation, grant 19-06489. KK acknowledges support from JSPS KAKENHI Grant Number JP17H06130. The work of YN was supported by NAOJ ALMA Scientific Research Grant Numbers 2017-06B and JSPS KAKENHI grant Number JP18K13577. TT acknowledges support from JSPS KAKENHI Grant Number JP20H00172. This paper makes use of the following ALMA data: ADS/JAO.ALMA\#2017.1.00586.S and ADS/JAO.ALMA\#2018.1.01684.S. ALMA is a partnership of ESO (representing its member states), NSF (USA) and NINS (Japan), together with NRC (Canada), MOST and ASIAA (Taiwan), and KASI (Republic of Korea), in cooperation with the Republic of Chile. The Joint ALMA Observatory is operated by ESO, AUI/NRAO and NAOJ.
This research has made use of the NASA/IPAC Extragalactic Database (NED), which is funded by the National Aeronautics and Space Administration and operated by the California Institute of Technology. The National Radio Astronomy Observatory is a facility of the National Science Foundation operated under cooperative agreement by Associated Universities, Inc. This research has made use of the SIMBAD database, operated at CDS, Strasbourg, France. Data analysis was in part carried out on the Multi-wavelength Data Analysis System operated by the Astronomy Data Center (ADC), National Astronomical Observatory of Japan.
\end{acknowledgments}

\vspace{5mm}

\software{
{\tt ALMA Calibration Pipeline},
{\tt Astropy} \citep{Astropy13,Astropy18},
{\tt CASA} \citep{McMullin07},
{\tt NumPy} \citep{Harris20},
{\tt PHANGS-ALMA Pipeline} \citep{Leroy21a},
{\tt SciPy} \citep{Virtanen20},
{\tt spectral-cube} \citep{Ginsburg19},
{\tt radio-beam}
}

\appendix
\section{Three biconical outflow models} \label{sec:appendix_a}
\begin{figure*}[ht!]
\begin{center}
\includegraphics[width=8.5cm]{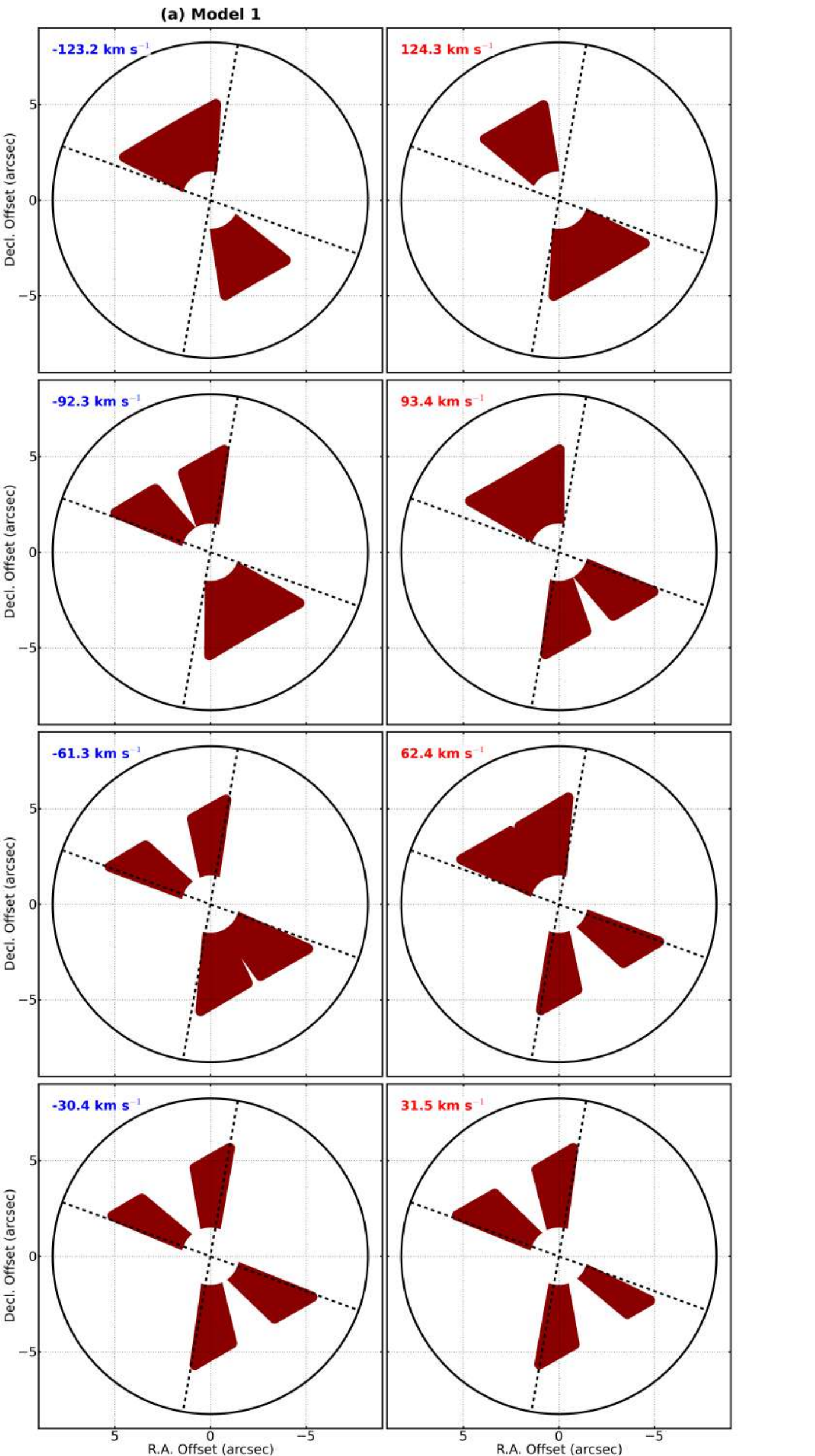}
\includegraphics[width=8.5cm]{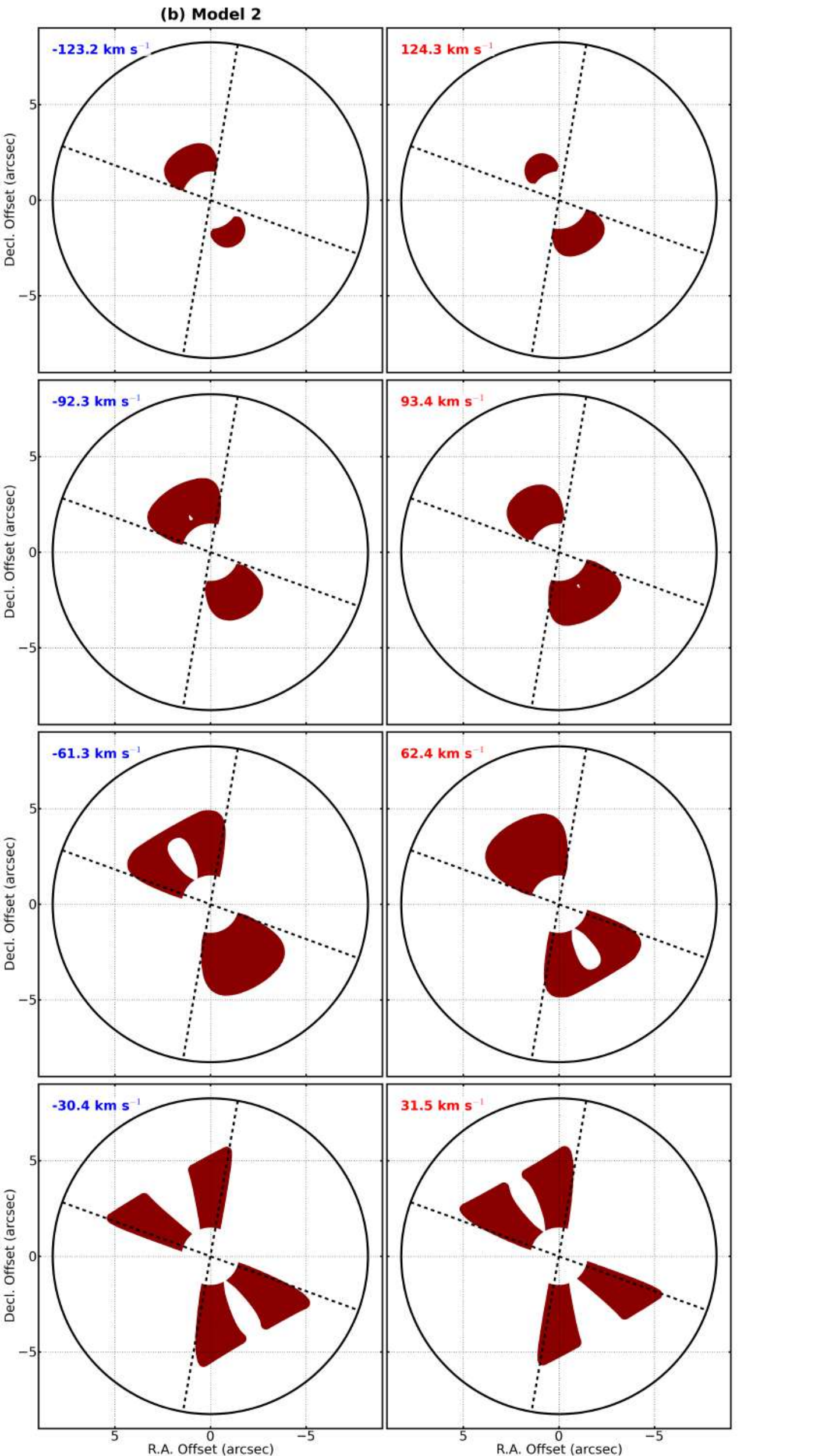}
\end{center}
\caption{
(a) The constant bicone model with no truncation (model 1). (b) The decelerating bicone model with no truncation (model 2).
\label{fig:channel_model}}
\end{figure*}

In Figure~\ref{fig:channel_model}, we show the two models (models 1 and 2) which do not reproduce the observed characteristics of the high \cishort/CO ratio gas in the central region of NGC~1068 (see Section~\ref{sec:modeling}). The observed channel maps and the channel maps of model 3 are shown in Figure~\ref{fig:channel}.

\section{Multiwavelength view of the outflow and the jet} \label{sec:appendix_b}
For comparison with our ALMA maps, we downloaded the MUSE \siii$\lambda\lambda$9069,9532/\sii$\lambda\lambda$6717,6731 ratio map\footnote{This is available at the CDS. See \url{http://cdsarc. u-strasbg.fr/viz-bin/qcat?J/A+A/622/A146}.}, proxy for the metallicity-independent ionization parameter (see \citealt{Mingozzi19} for a detailed description). In addition, the 8.49~GHz radio continuum map taken by Karl G. Jansky Very Large Array (VLA) is retrieved from the VLA Data Archive\footnote{\url{https://archive.nrao.edu/archive}}. We also downloaded
drizzled {\it HST} WFPC2 F502N and F547M maps from the Hubble Legacy Archive\footnote{\url{https://hla.stsci.edu/}} in order to create a \oiii$\lambda\lambda$5007 map. Both the VLA and {\it HST} maps have higher angular resolution and image a larger area than the \ci\ map.

\begin{figure*}[ht!]
\begin{center}
\includegraphics[width=15.5cm]{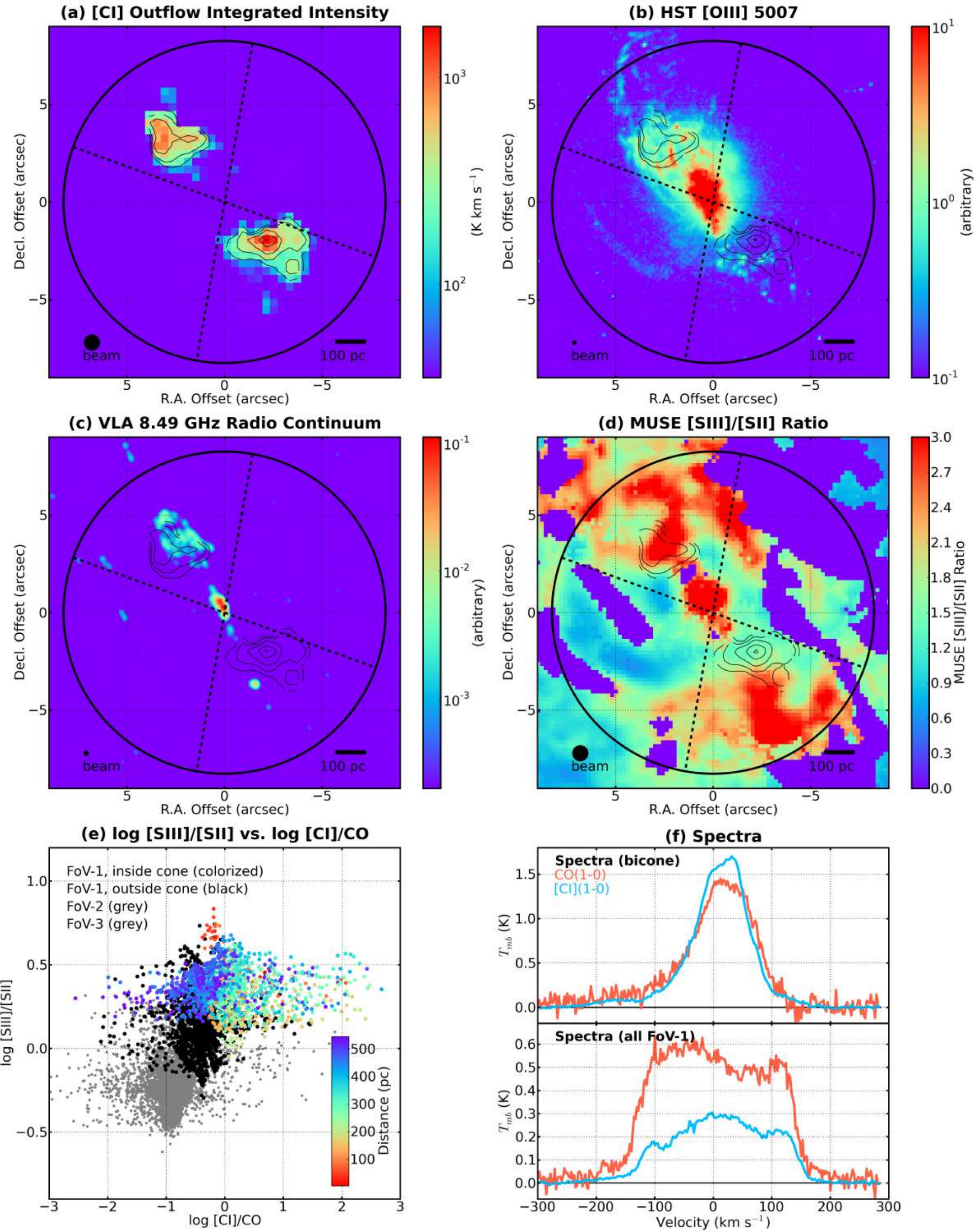}
\end{center}
\caption{
(a) The \cishort\ outflow intensity map (see Appendix~\ref{sec:appendix_b}). (b) {\it HST} \oiii\ map. (c) VLA 8.49~GHz continuum map. (d) VLT/MUSE \siii/\sii\ line ratio map (i.e., ionization parameter map). (e) Pixel-by-pixel comparison between the \ci/\coone\ ratio and \siii/\sii\ ratio maps. Pixels within the cones in FoV 1 are color-coded by projected distance from the AGN, and pixels lying outside the cones in FoV 1 are shown in black. Pixels in FoV 2 and FoV 3 (i.e., starburst ring) are shown in grey. (f) \ci\ and \coone\ spectra toward the bicone (top) and FoV-1 (bottom). We use the \cishort\ detected pixels of Figure~\ref{fig:multi}a when extracting the bicone spectra.
\label{fig:multi}}
\end{figure*}

In Figure~\ref{fig:multi}b-d, those three ancillary maps are shown, as well as our \cishort\ map for comparison. The \cishort\ outflow intensity map (Figure~\ref{fig:multi}a) is created as follows: (1) create a mask cube using the line ratio cube (Figure~\ref{fig:channel}a) clipped at ratio $>$ 1, (2) apply the mask to the \cishort\ datacube, and then (3) collapse it. We note that the \cishort\ intensity map shown in Figure~\ref{fig:map}a is created without this line ratio mask.

In Figure~\ref{fig:multi}e, we show a pixel-by-pixel comparison between the \siii/\sii\ ratio and \cishort/CO ratio maps in the same manner as Figure~\ref{fig:map}d. Data points within the bicone of FoV-1 show systematically higher ratio values compared to the data from FoV-2 and FoV-3 (i.e., starburst ring), implying that normal star-forming activities happening in NGC~1068 cannot explain the observed values within the bicone.

We show averaged CO and \cishort\ spectra in Figure~\ref{fig:multi}f. As, in general, molecular outflow studies focus on line profiles, it is important to check how the \cishort\ outflow spectrum looks. To make outflow spectra, we employ a fine-tuned version of the method described in \citet{Stuber21}, i.e., extracting outflow spectra based on a mask and comparing it with a galaxy-averaged spectrum.

\section{Derived parameters of the cold gas outflow} \label{sec:appendix_c}
Here we briefly summarize the parameters of the best bicone model (i.e., model 3) for the \cishort\ outflow. As we basically applied the D06 model that well describes the \oiii\ outflow of NGC~1068, we use the same symbols as defined in D06.

The maximum distance along the bicone axis from the nucleus $z_{\rm max}$ is 400~pc, the inner opening angle of the bicone $\theta_{\rm inner}$ is 20$\degr$, the outer opening angle $\theta_{\rm outer}$ is 40$\degr$, the inclination angle between the bocine axis and the plane of the sky $i_{\rm axis}$ is 5$\degr$, and the position angle PA$_{\rm axis}$ is 30$\degr$. These values are exactly the same as the D06 model, and we did not need to modify these parameters to explain the \cishort\ and CO data. However, we need a more sophisticated model to constrain them.

We also assumed the velocity structure employed by the D06 model, i.e., $v_{\rm int}(r) = v_{\rm max}-k(r-r_t)$ (see Section~\ref{sec:modeling}). This decelerating model is required to explain the observed characteristics of the \cishort\ and CO outflows, although the maximum velocity of the \oiii\ outflow ($v_{\rm max}$ $\sim$ 2000~\kms) is too fast. After some exploration of the parameter space from 100\ to 2000\kms, we found that $v_{\rm max}$ $\sim$ 300~\kms\ can reproduce the observed \cishort\ and CO channel maps.

We suggest that the bright \cishort\ emission comes from the region where the ionized gas outflow hits the galactic disk in order to explain the ``imperfect" bicone geometry. We employed the disk model described in \citet{Schinnerer00}. This spatial configuration is consistent with the model described in \citet{Garcia-Burillo14,Garcia-Burillo19}. Based on the configuration, there are two regions (north and south) where \cishort\ emission is enhanced as shown in Figure~\ref{fig:multi}a. The measured \cishort\ luminosity and CO luminosity of the northern outflow are 10$^{5.37}$~\Kkms~pc$^2$ and 10$^{5.33}$~\Kkms~pc$^2$, and the luminosities of the southern outflow are 10$^{5.41}$~\Kkms~pc$^2$ and 10$^{5.25}$~\Kkms~pc$^2$, respectively. Although measuring molecular gas masses is important to characterise the cold gas outflow of this galaxy, it requires additional datasets and efforts (e.g., need a gold standard gas mass tracer). This is an interesting topic for future projects.

\bibliography{n1068_ci_outflow}{}
\bibliographystyle{aasjournal}

\end{document}